\documentclass[lettersize,journal]{IEEEtran}
\usepackage{amsmath,amsfonts}
\usepackage{array}
\usepackage{subfig}
\usepackage{stfloats}
\usepackage{url}
\usepackage{verbatim}
\usepackage{graphicx}
\usepackage{cite}
\usepackage{microtype}
\usepackage{doi}
\usepackage{float}
\usepackage{mathrsfs}
\usepackage{threeparttable}
\usepackage{booktabs}
\usepackage{multicol}
\usepackage{multirow}
\usepackage{tabto}
\usepackage[ruled]{algorithm2e}
\usepackage{algpseudocode}
\usepackage{makecell}
\usepackage{array}
\usepackage{diagbox}
\usepackage{adjustbox}
\usepackage{hyperref}
\usepackage{textcomp}
\usepackage{stackrel,amssymb}
\usepackage{xcolor}

\begin{document}
 \title{Diffusion-based Counterfactual Augmentation: Towards Robust and Interpretable Knee Osteoarthritis Grading}
 
\author{Zhe Wang, Yuhua Ru, Aladine Chetouani, Tina Shiang, Fang Chen, Fabian Bauer, Liping Zhang, Didier Hans, Rachid Jennane, William Ewing Palmer$^*$, Mohamed Jarraya$^*$, Yung Hsin Chen$^*$
\thanks{Zhe Wang, Mohamed Jarraya, William Ewing Palmer, Tina Shiang, and Yung Hsin Chen are with Department of Radiology, Harvard Medical School, Boston, 02114, USA (e-mails: zwang78@mgh.harvard.edu; mjarraya@mgh.harvard.edu; wpalmer@mgh.harvard.edu; tshiang@bwh.harvard.edu; ychen4@mgh.harvard.edu).}
\thanks{Yuhua Ru is with Jiangsu Institute of Hematology, The First Affiliated Hospital of Soochow University, Suzhou, 215006, China (e-mail: ruyuhua@suda.edu.cn).}
\thanks{Aladine Chetouani is with L2TI Laboratory, University Sorbonne Paris Nord, Villetaneuse, 93430, France (e-mail: aladine.chetouani@univ-paris13.fr).}
\thanks{Fang Chen is with department of Medical School, Henan University of Chinese Medicine, Zhengzhou, 450046, China (e-mail: chenfangyxy@hactcm.edu.cn).}
\thanks{Fabian Bauer is with Institute for Diagnostic and Interventional Radiology, Faculty of Medicine and University Hospital Cologne, University of Cologne, 50937, Cologne, Germany (e-mail: fabian.bauer@uk-koeln.de).}
\thanks{Liping Zhang is with Department of Electrical and Electronic Engineering, University of Hong Kong, 999077, Pokfulam, Hong Kong (e-mail: lpzhang@link.cuhk.edu.hk).}
\thanks{Didier Hans is with Nuclear Medicine Division, Geneva University Hospital, Geneva, 1205, Switzerland (e-mail: didier.hans@chuv.ch).}
\thanks{Rachid Jennane is with IDP Institute, UMR CNRS 7013, University of Orleans, Orleans, 45067, France (e-mail: rachid.jennane@univ-orleans.fr).}}


\maketitle
\begin{abstract}
Automated grading of Knee Osteoarthritis (KOA) from radiographs is challenged by significant inter-observer variability and the limited robustness of deep learning models, particularly near critical decision boundaries. To address these limitations, this paper proposes a novel framework, Diffusion-based Counterfactual Augmentation (DCA), which enhances model robustness and interpretability by generating targeted counterfactual examples. The method navigates the latent space of a diffusion model using a Stochastic Differential Equation (SDE), governed by balancing a classifier-informed boundary drive with a manifold constraint. The resulting counterfactuals are then used within a self-corrective learning strategy to improve the classifier by focusing on its specific areas of uncertainty. Extensive experiments on the public Osteoarthritis Initiative (OAI) and Multicenter Osteoarthritis Study (MOST) datasets demonstrate that this approach significantly improves classification accuracy across multiple model architectures. Furthermore, the method provides interpretability by visualizing minimal pathological changes and revealing that the learned latent space topology aligns with clinical knowledge of KOA progression. The DCA framework effectively converts model uncertainty into a robust training signal, offering a promising pathway to developing more accurate and trustworthy automated diagnostic systems. Our code is available at \url{https://github.com/ZWang78/DCA}.
\end{abstract}


\section{Introduction}
\IEEEPARstart{K}{nee} OsteoArthritis (KOA) is a prevalent chronic degenerative joint disease characterized by the progressive deterioration of cartilage, Joint Space Narrowing (JSN), osteophyte formation, and subchondral bone remodelling \cite{kneeoa}. It significantly impairs mobility, reduces quality of life, and represents a considerable burden on healthcare systems globally \cite{multi-factor}. The prevalence of KOA is expected to rise substantially due to an ageing population, increased obesity rates, and heightened physical activity, thereby amplifying the societal and economic impacts \cite{nocure}. Radiographic evaluation remains the clinical gold standard for assessing KOA severity \cite{kohn2016classifications}, primarily utilizing the Kellgren–Lawrence (KL) grading system \cite{KL}, which categorizes disease progression into five ordinal stages, shown in Table \ref{KL_grades}. Despite its widespread usage and acceptance, KL grading is limited by inter-observer variability \cite{yassine}. This variability arises from subtle and often ambiguous anatomical distinctions visible on X-ray images, compounded by factors such as image acquisition quality, radiographic positioning variations, and the subjective interpretation of visual cues among clinicians \cite{chen}. Consequently, diagnostic uncertainty persists, hindering reliability and potentially impacting clinical decision-making, patient management, and therapeutic outcomes \cite{zhe}.

\begin{table}[htbp]
\centering
\caption{Description of the KL grading system for KOA}
\setlength{\tabcolsep}{2mm}
\begin{tabular}{lll}
\toprule
Grade &  Severity & Description\\
\midrule
KL-0 & none & none of osteoarthritis\\
KL-1 & doubtful & potential osteophytic lipping \\
KL-2 & minimal & certain osteophytes and potential JSN\\
KL-3 & moderate & moderate multiple osteophytes, certain JSN,\\
& & and some sclerosis\\
KL-4 & severe & large osteophytes, certain JSN, and severe sclerosis\\
\bottomrule
\end{tabular}
\label{KL_grades}
\end{table}

Deep learning techniques have now substantially advanced automated medical image analysis, demonstrating promising performance in various diagnostic tasks, including KOA severity grading \cite{greenspan2016guest}. Convolutional Neural Networks (CNNs) \cite{cnn} and transformer-based \cite{vaswani2017attention} architectures have been extensively explored, achieving 
encouraging performance in detecting subtle radiographic features associated with KOA progression \cite{antony2016} \cite{antony1} \cite{tiulpin2} \cite{tiuplin} \cite{wang2024transformer}. Despite these successes, learning-based models typically exhibit limited robustness and generalization \cite{schmidt2018adversarially}, especially near critical decision boundaries \cite{lei2023understanding}, where minor structural variations may lead to significant fluctuations in predicted KL grades, which poses substantial challenges to clinical integration, as reliable predictions near borderline cases are crucial for effective patient management \cite{dawson2013relationship}. While traditional data augmentation techniques, such as geometric transformations including flipping, rotation, and scaling, are widely employed to increase dataset diversity to improve the robustness of the models, they remain insufficient for capturing the clinically nuanced and pathologically relevant variations observed between adjacent stages \cite{heavily_rely}. Moreover, current deep learning models frequently operate as "black boxes", providing minimal interpretability regarding the structural features influencing their classification decisions \cite{guidotti2018survey}. Therefore, enhancing model interpretability alongside robustness has become an essential focus to facilitate broader clinical adoption and ensure meaningful integration into diagnostic workflows \cite{castaneda2015clinical}.

\section{Related work}

Methodological advancements in data augmentation have generated more diverse and clinically realistic training examples, significantly enhancing the performance and generalizability of automated medical imaging analysis by enriching training datasets. For instance, in \cite{frid2018gan}, Frid-Adar et al. utilized Deep Convolutional Generative Adversarial Networks (DCGANs) to generate synthetic yet realistic Computed Tomography (CT) images of liver lesions. The results showed that augmenting a classifier's training set with these synthetic images, in addition to traditional augmentation, significantly improved diagnostic performance. Similarly, in \cite{calimeri2017biomedical}, Calimeri et al. used GANs to address the common challenge of data scarcity in Alzheimer's disease research. The authors generated synthetic brain MRI scans to enhance the generalization capability of a CNN model tasked with distinguishing between healthy controls and patients with Alzheimer's disease. Back to our objective, KL grade-specific augmentation strategies have emerged to target the challenging differentiation between adjacent grades.  In \cite{prezja2022deepfake}, Prezja et al. developed two Wasserstein GANs trained on 5,556 real knee X-rays to generate 320,000 synthetic radiographs across different KL grades. These images were used to augment KL classification tasks, achieving a notable performance gain. In \cite{wang2023key}, Wang et al. proposed Key-Exchange Convolutional Auto-Encoder (KECAE) designed to enhance early KOA detection, performing structured semantic augmentation by selectively exchanging key pathological features between KL-0 and KL-2 X-rays in latent space. Experiments show that KECAE-augmented data consistently improves classification accuracy across various architectures. Furthermore, in \cite{wang2024temporal}, Wang et al. proposed a Diffusion-based Morphing Model (DMM) to simulate the temporal evolution of KOA from KL-0 to KL-4 using pairs of radiographs from the same patient. This method integrates a Denoising Diffusion Probabilistic Model (DDPM) \cite{ddpm} with a registration-based morphing module to generate anatomically plausible intermediate frames between healthy and severely degenerated knees. Experimental results show that augmenting classification models with the generated frames significantly improved early-stage KOA classification accuracy.

Recent advancements in counterfactual generation methods show significant potential for enhancing medical imaging analysis by offering interpretable insights and generating targeted augmentation data. For instance, in \cite{atad2024counterfactual}, a framework by Atad et al. uses a Diffusion AutoEncoder (DAE) to produce counterfactuals directly from the model's latent space to model the continuous, ordinal progression of pathologies such as vertebral compression fractures and diabetic retinopathy. By manipulating latent codes, the model synthesizes visual explanations that illustrate the minimal changes needed to alter a diagnosis. In \cite{ye2023pairwise}, Ye et al. developed an adversarial framework for Alzheimer's classification where a generator and classifier iteratively identify the hardest counterfactuals to improve model robustness and data efficiency. Similarly, in \cite{xia2024mitigating}, Xia et al. employed soft counterfactual fine-tuning to mitigate attribute amplification to ensure causal fidelity in chest X-ray analysis, preventing spurious alterations to clinical signs when modifying protected attributes and thereby producing more trustworthy explanations for clinical deployment. Despite the promising advancements in recent augmentation strategies, several significant limitations persist. Firstly, most existing methods generate augmented samples without explicitly considering the classifier’s decision boundaries. Although GAN- and autoencoder-based approaches can effectively synthesize images, they typically lack direct sensitivity to regions where the classifier exhibits heightened uncertainty, particularly around critical class transitions. Additionally, current methodologies often treat the processes of classifier training and data augmentation as independent stages, thus failing to dynamically adapt augmentation strategies to target areas where the classifier is most vulnerable. This research gap is especially evident in domains with fine-grained features. To the best of our knowledge, no efficient counterfactual approaches have yet been applied to KOA research. To address these critical limitations, we propose Diffusion-based Counterfactual Augmentation (DCA), a novel framework that seamlessly integrates semantic perturbations with explicit sensitivity to decision boundaries, to generate pathologically coherent counterfactuals that populate sparsely sampled regions near the classifier's margin, thereby creating more robust and balanced training set, facilitating more robust training and improved interpretability in KOA grading.

The primary contributions of this study include the following:
\begin{itemize}
\item[$\bullet$] \textbf{Counterfactual generation:} We introduce a novel gradient-informed counterfactual generation framework that explicitly targets classifier decision boundaries in the latent space. By balancing a classifier-driven boundary force with a learned manifold constraint, the proposed method generates semantically meaningful radiographs that reflect subtle pathological transitions between adjacent KOA grades.
\item[$\bullet$] \textbf{Self-corrective learning:} A pair of structurally identical classifiers, a frozen reference and a trainable counterpart, work in tandem within a self-corrective learning loop. The static classifier guides counterfactual generation by identifying uncertain boundary regions, while the learnable classifier is trained on the generated samples to resolve prediction ambiguities and progressively improve robustness near decision boundaries.
\item[$\bullet$] \textbf{Comprehensive validation:} We conduct extensive experiments on two large-scale, real-world KOA cohorts, evaluating the proposed framework across multiple classification architectures. The results demonstrate consistent improvements in accuracy, generalizability, and interpretability, supported by both quantitative metrics and qualitative visualization analyses.
\end{itemize}

\section{Methodology}
\begin{figure*}[htbp]
\centering 
\includegraphics[width=1\textwidth]{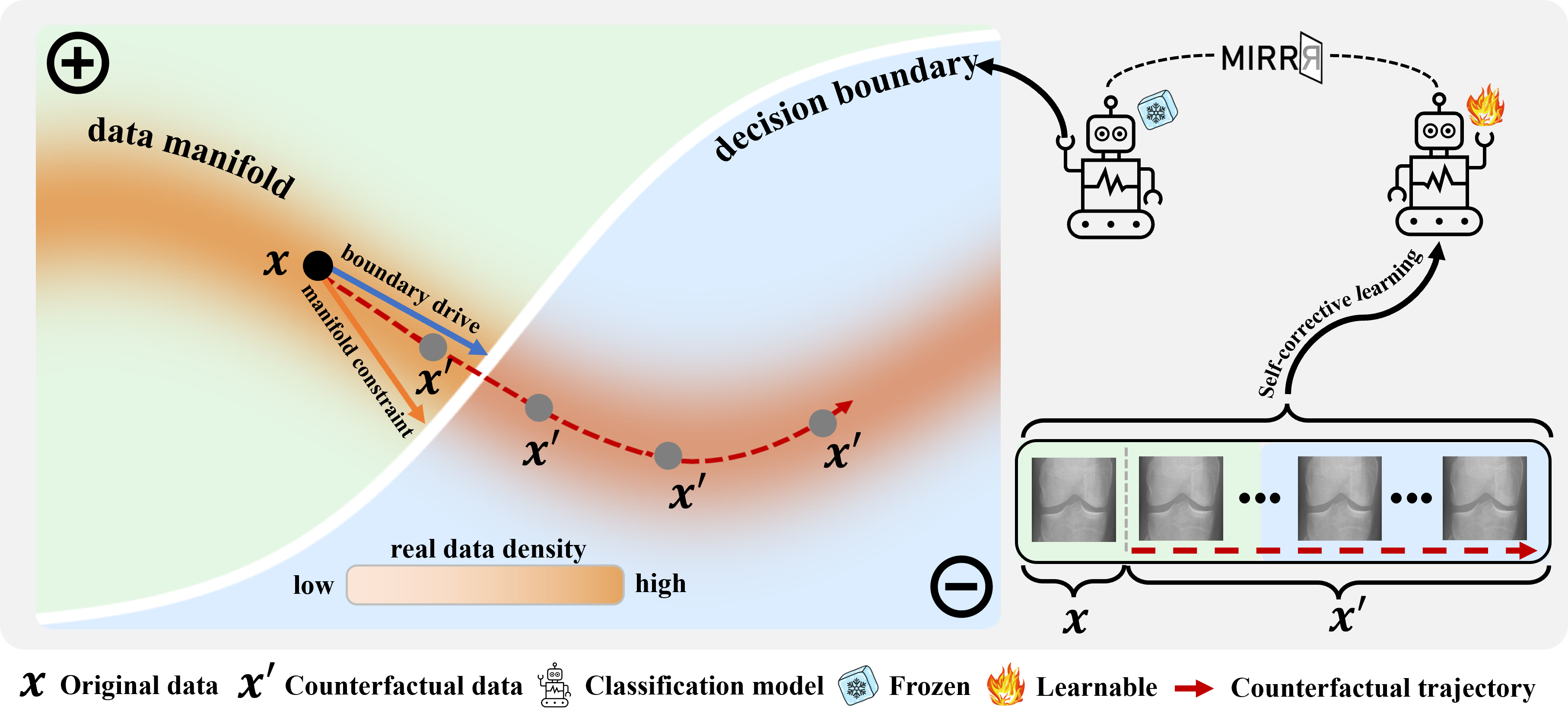}
\caption{The flowchart of manifold-constrained counterfactual data generation. Starting from an original data point $x$ residing on the data manifold within its source class region (+), a trajectory (dashed red line) towards the target class region (-) is generated to produce counterfactual samples $x'$. This trajectory is guided by balancing two competing influences: (1) a boundary drive (blue arrow), derived from a parameter-frozen classification model, oriented towards the decision boundary defined by this frozen classifier to guide the data across it into the decision boundary towards the target class region, and (2) a manifold constraint (orange arrow), derived from the learned data distribution, pulling the original data towards the highest-density regions of the data manifold to maintain realism. All original ($x$) and generated counterfactual data ($x'$) serve as augmented data for self-corrective learning to improve the performance and robustness of classification models.}
\label{flowchart}
\end{figure*}

An overview of the framework is illustrated in Fig. \ref{flowchart}. The method initiates a semantic trajectory from an original data point, $x$, evolving it towards a target class region. By sampling various intermediate data points, $x'$, along this path, the framework produces a rich set of counterfactuals that map the space between the two classes. These generated counterfactuals ultimately serve as augmented data for a self-corrective learning strategy to improve the classification model.


\subsection{Semantic-factored boundary exploration}
The proposed framework commences by leveraging a Latent Diffusion Model (LDM) \cite{latent_diffusion_model}, pre-trained on knee radiograph datasets, to obtain compact and semantically meaningful latent representations. The encoder of the LDM, denoted by $\mathcal{E}$, maps each input radiograph $x$ to a latent representation $z$, serves as the starting point for the exploration:

\begin{equation}
z = \mathcal{E}(x), \quad z \in \mathbb{R}^{n \times d}
\label{encoder}
\end{equation}
where $n$ signifies the spatial dimensionality of the latent embedding, while $d$ represents the channel dimension.

To systematically explore the transition boundaries between adjacent KL grades, we employ a Stochastic Differential Equation (SDE) \cite{kloeden1992stochastic} to simulate a guided trajectory in the latent space. Starting from an initial latent state $z^{(0)}$ corresponding to the source grade $y$, the trajectory evolves toward the decision boundary of a target grade $y^\prime$. This evolution is modelled over a continuous time horizon $s \in [0, T_{\text{SDE}}]$, where the latent state at time $s$ is denoted by $z^{(s)}$, with $z^{(0)}$ as the initial condition. The SDE governs the infinitesimal change in state $dz$ as follows:

  
\begin{equation}
dz = f(z^{(s)})\,ds + \gamma(s)\,dw \label{eq:sde_full}
\end{equation}
where $w$ represents a standard Wiener process \cite{shale1968wiener}, and the term $\gamma(s) = 1 - s/T_{\text{SDE}}$ is a time-dependent diffusion coefficient. The drift term $f(z^{(s)})$ provides semantic guidance and is defined as a weighted combination of two directional forces:

\begin{equation}
f(z^{(s)}) = (1 - \lambda) \cdot \mathbf{v}_{\text{manifold}}^{(s)} + \lambda \cdot \mathbf{v}_{\text{boundary}}^{(s)}
\end{equation}
where $\lambda \in [0, 1]$ is a hyper-parameter controlling the balance between manifold adherence and semantic guidance. 

The first term, $\mathbf{v}_{\text{manifold}}^{(s)}$, encourages the trajectory to remain within high-density regions of the learned data distribution. It is derived from the score function estimated using a denoising network $\epsilon_\theta$ from the diffusion model:


\begin{equation}
\mathbf{v}_{\text{manifold}}^{(s)} = \frac{\nabla_{z^{(s)}} \log p(z^{(s)})}{||\nabla_{z^{(s)}} \log p(z^{(s)})||_2}
\end{equation}
where 
\begin{equation}
\nabla_{z^{(s)}} \log p(z^{(s)}) \approx -\frac{\epsilon_\theta(z^{(s)}, t)}{\sqrt{1-\bar{\alpha}_t}}
\end{equation}
where $\bar{\alpha}_t = \prod_{i=1}^t \alpha_i$ is the cumulative noise schedule parameter at timestep $t$.

The second term, $\mathbf{v}_{\text{boundary}}^{(s)}$, introduces directional movement along semantic axes. We compute a raw gradient signal indicating the direction towards the target KL grade boundary in the image space $x^{(s)}=\mathcal{D}(z^{(s)})$. This direction is computed directly from the gradient of a frozen reference classifier $\mathcal{C}^*$:



\begin{equation}
\mathbf{v}_{\text{boundary}}^{(s)} = \frac{g^{(s)}}{||g^{(s)}||_2}
\end{equation}
where
\begin{equation}
g^{(s)} = \nabla_{x^{(s)}} \left( \log P(y'|x^{(s)}) - \log P(y|x^{(s)}) \right)
\end{equation}



Finally, the continuous SDE (Eq. \ref{eq:sde_full}) is simulated numerically using the Euler-Maruyama discretization scheme \cite{kloeden1992stochastic}. The latent state is updated iteratively as:

\begin{equation}
z^{(s+\Delta s)} = \underbrace{z^{(s)}}_{\text{current state}}
+
\underbrace{f(z^{(s)})\Delta s}_{\text{semantic guidance}}
+
\underbrace{\gamma(s)\sqrt{\Delta s} \cdot \epsilon}_{\text{random perturbation}} \label{eq:euler_maruyama}
\end{equation}
where $\Delta s = T_{\text{SDE}} / N_{\text{SDE}}$ is the discretized step size over $N_{\text{SDE}}$ total steps, and $\epsilon \sim \mathcal{N}(0, I)$ is Gaussian noise sampled independently at each iteration.

\subsection{Diffusion-based refinement}
Although the classifier-guided SDE provides a powerful mechanism for traversing counterfactual trajectories across KL grade boundaries, the latent codes it generates may still exhibit subtle artefacts or deviate from the true data manifold of anatomically plausible structures. These imperfections are primarily due to the intrinsic stochasticity of SDE-based sampling. To address this, we introduce a diffusion-based refinement stage designed to enhance anatomical realism and semantic fidelity. Specifically, this stage applies a limited number of reverse diffusion steps, the last 50 steps in our implementation, to realign the generated latent codes with the learned distribution of realistic radiographic features. It is noteworthy that such a refinement process is not restricted to the final latent state $z^{(T_{\text{SDE}})}$. Instead, it is applied to the full set of intermediate latent states ${z^{(s)}}$ sampled along the SDE trajectory. These intermediate representations are individually refined using a reverse Markov chain process \cite{norris1998markov}, governed by the standard denoising formulation of the diffusion model:

\begin{equation}
\label{eq:refinement_ddpm}
\begin{aligned}
z_{t-1} =\ 
&\frac{1}{\sqrt{\alpha_t}} \left( 
z_t - \frac{1 - \alpha_t}{\sqrt{1 - \bar{\alpha}_t}} \cdot \epsilon_\theta(z_t, t) 
\right) \\
&+ \sqrt{\frac{(1 - \alpha_t)(1 - \bar{\alpha}_{t-1})}{1 - \bar{\alpha}_t}} \cdot \epsilon,\quad 
\epsilon \sim \mathcal{N}(0, I)
\end{aligned}
\end{equation}
where $\alpha_t$ and $\bar{\alpha}_t = \prod_{i=1}^t \alpha_i$ are parameters from the noise schedule, and $\epsilon$ denotes standard Gaussian noise.

\subsection{Self-corrective learning}
Following the SDE-guided exploration and diffusion-based refinement, we obtain a set of high-fidelity counterfactual samples $x'$, whose predicted class shifts from a source label $y$ to a target label $y'$. While these samples may not lie exactly on the decision boundary, they inherently traverse regions of elevated model uncertainty, making them highly informative for targeted augmentation. To leverage this information, we introduce a self-corrective learning strategy involving two classifiers with identical architectures: a frozen reference classifier $\mathcal{C}^*$ and a learnable target classifier $\mathcal{C}$. The core objective is to improve the robustness of $\mathcal{C}$ by exposing it to challenging, boundary-adjacent samples and guiding its predictions using the stable outputs of $\mathcal{C}^*$. It is noteworthy that although the reference classifier $\mathcal{C}^*$ may have its own inherent limitations, its frozen nature provides a temporally consistent decision function, acting as a reliable anchor in regions where the learnable classifier $\mathcal{C}$ is likely to behave erratically. By enforcing local prediction consistency between $\mathcal{C}$ and $\mathcal{C}^*$ on the counterfactual inputs $x'$, we regularize $\mathcal{C}$ in semantically ambiguous regions—without impeding its ability to generalize on the original data distribution.

To this end, we adopt a multi-task loss with learnable uncertainty weights, formulated as:

\begin{equation}
\mathcal{L}_{\text{total}} = \frac{1}{\sigma_{\text{CE}}^2} \mathcal{L}_{\text{CE}} + \frac{1}{\sigma_{\text{align}}^2} \mathcal{L}_{\text{align}} + \log \sigma_{\text{CE}} + \log \sigma_{\text{align}} \label{eq:hybrid_loss}
\end{equation}
where $\mathcal{L}_{\text{CE}}$ is the conventional cross-entropy loss, applied to both original samples $(x, y)$ and counterfactual samples $(x', y')$. The terms $\sigma_{\text{CE}}$ and $\sigma_{\text{align}}$ are learnable parameters representing the homoscedastic uncertainty associated with each loss component. The corresponding logarithmic terms serve as regularizers. The alignment loss $\mathcal{L}_{\text{align}}$ encourages the learnable classifier $\mathcal{C}$ to mimic the predictions of $\mathcal{C}^*$ on counterfactual samples:

\begin{equation}
\mathcal{L}_{\text{align}} = \left\| \mathcal{C}(x') - \mathcal{C}^*(x') \right\|_2^2 
\label{eq:align_loss}
\end{equation}


\section{Experimental data and settings}

\subsection{Employed database}
The empirical validation of our proposed methodology leverages two publicly accessible longitudinal cohorts renowned for research in KOA: the Osteoarthritis Initiative (OAI) \cite{OAI} and the Multicenter Osteoarthritis Study (MOST) \cite{segal2013multicenter}, specifically:

\subsubsection{OAI}
The OAI is a multi-center, longitudinal study involving 4,796 participants aged 45 to 79, at varying risk levels for developing KOA. Initiated in 2002 with support from the National Institutes of Health (NIH) and private sector partners, the OAI was designed to identify and evaluate biomarkers for KOA development and progression.

\subsubsection{MOST}
To further assess the robustness and generalizability of our approach, we incorporated data from the MOST, which is a complementary large-scale, longitudinal cohort study, enrolling 3,026 participants aged 50 to 79 years. It focuses on elucidating risk factors for incident and progressive KOA, particularly biomechanical, genetic, and lifestyle elements, with radiographic data collected at baseline and multiple follow-up intervals.

Specifically, Table \ref{data_distribution} details the specific number of cases available for each KL grade within both the OAI and MOST cohorts utilized in this study. 

\begin{table}[htbp]
    \centering
    \caption{Data distribution of the employed databases}
    \setlength{\tabcolsep}{3.5mm}
    \begin{tabular}{lccccc}
        \toprule
        Database & KL-0 & KL-1 & KL-2 & KL-3 & KL-4 \\
        \midrule
        OAI & 3,388 & 1,597 & 2,374 & 1,239 & 235 \\
        MOST & 2,521 & 1,028 & 936 & 1,001 & 460\\
        \bottomrule
    \end{tabular}
    \label{data_distribution}
\end{table}


\subsection{Experimental details}
The generative framework consists of a $\beta$-Variational Autoencoder ($\beta$-VAE) \cite{burgess2018understanding}, a pre-trained classifier, and a diffusion U-Net. The $\beta$-VAE was configured with an encoder-decoder architecture with channel dimensions of 32, 64, 128, 256, and 512, using a single residual block per resolution level. The latent space is defined with 32 channels. For classification guidance, a Vision Transformer-based classifier from \cite{wang2024transformer}, pre-trained on the target KOA datasets, was utilized. The diffusion U-Net model was designed with channel sizes of 128, 256, 512, and 1024, incorporating two residual blocks per resolution and self-attention modules at spatial resolutions of 14$\times$14 and 28$\times$28. The temporal embedding dimension is set to 128. Both the $\beta$-VAE and diffusion U-Net are trained using the Adam optimizer \cite{adam}. The $\beta$-VAE is trained for 2,000 epochs with a learning rate of $2 \times 10^{-5}$ and a $\beta$ value of 0.9. The U-Net is subsequently trained for 50,000 iterations using a learning rate of $1 \times 10^{-4}$. A consistent batch size of 64 is used for all training procedures. The diffusion process is configured with $T=200$ timesteps and a linear noise schedule. Counterfactual samples are generated via an SDE solver, executed over 1,000 steps with a classifier guidance scale of 2.5. This is followed by a 50-step guided diffusion refinement phase. All implementations are developed in PyTorch v1.12.1 \cite{pytorch} and executed on NVIDIA A100 GPUs with 80 GB memory.

\section{Experiments and results}

\subsection{Ablation study for diffusion-based refinement}
To assess the necessity and effectiveness of the diffusion-based counterfactual refinement module, we conducted an ablation study that evaluated the impact of removing this component. Specifically, we compared performance between models that included the refinement stage ("W/") and those that used raw SDE-generated latent codes without denoising ("W/O"). Two quantitative metrics were employed: Maximum Mean Discrepancy (MMD, scaled by $10^{-2}$) to measure distributional dissimilarity between generated counterfactuals and original data, and average k-Nearest Neighbor (k-NN) distances to assess the fidelity of individual samples by evaluating their proximity to real data points. For this analysis, counterfactuals were generated for each target KL grade from its adjacent grade, with KL-0 and KL-4 synthesized unidirectionally from KL-1 and KL-3, respectively. All metrics were averaged over three independent generation trials. As reported in Table \ref{mmd_knn_results}, the model incorporating diffusion refinement consistently achieved lower MMD and k-NN distances across all KL grades, indicating that the refinement stage effectively enhances anatomical plausibility by constraining the generated samples to remain closer to the true data manifold. Furthermore, an observable trend in the results is a general increase in both MMD and k-NN distances as the KL grade advances, which can be attributed to the fact that, as KOA progresses, the structural changes become more complex, necessitating a higher degree of precision in the generated counterfactuals to accurately capture these pathological features and maintain proximity to the true data manifold.

\begin{table}[htbp]
\centering
\scriptsize
\setlength{\tabcolsep}{1.3mm}
\caption{Comparison of the generated counterfactual and original KL-X samples on the OAI database}
\label{mmd_knn_results}
\begin{threeparttable}
\begin{tabular}{llll}
\toprule
Target & Diff & MMD & Average 1-NN / 5-NN / 10-NN distance \\
\midrule
\multirow{2}{*}{KL-0} & W/& $4.82\pm1.71$ & $1.24\pm0.13$ /  $1.54\pm0.65$ / $1.61\pm0.23$\\
 & W/O & $18.32\pm5.93$ & $4.72\pm0.46$ / $5.60\pm0.71$ / $5.91\pm0.92$\\
\midrule
\multirow{2}{*}{KL-1} & W/& $5.27\pm1.81$ & $1.52\pm0.21$ / $1.92\pm0.40$ / $2.42\pm0.37$\\
& W/O & $27.29\pm9.25$ & $5.22\pm0.58$ / $5.94\pm0.47$ / $6.28\pm0.56$ \\
\midrule
\multirow{2}{*}{KL-2} & W/& $6.12\pm1.94$ & $1.48\pm0.79$ / $1.99\pm0.61$ / $2.12\pm0.63$ \\
 & W/O & $22.12\pm5.78$ & $9.61\pm0.74$ / $10.39\pm0.95$ / $10.95\pm1.17$\\
\midrule
\multirow{2}{*}{KL-3} & W/& $9.21\pm1.89$ & $1.77\pm0.15$ / $2.58\pm0.89$ / $2.85\pm0.66$ \\
 & W/O & $31.93\pm6.01$ & $9.56\pm0.80$ / $10.44\pm0.73$ / $11.38\pm0.68$\\
\midrule
\multirow{2}{*}{KL-4} & W/& $7.07\pm1.05$ & $2.29\pm0.57$ / $2.62\pm0.55$ / $2.88\pm0.64$\\
& W/O & $30.57\pm8.51$ & $8.97\pm0.93$ / $9.68\pm0.88$ / $10.20\pm0.72$\\
\bottomrule
\end{tabular}
\begin{tablenotes}
\footnotesize
\item[$*$] MMD was computed using a Radial Basis Function (RBF) \cite{vert2004primer} kernel. 
\item[] Average k-NN distances were computed using the Euclidean metric.
\end{tablenotes}
\end{threeparttable}
\end{table}

\begin{figure}[htbp]
\centering
\includegraphics[width=0.4\textwidth]{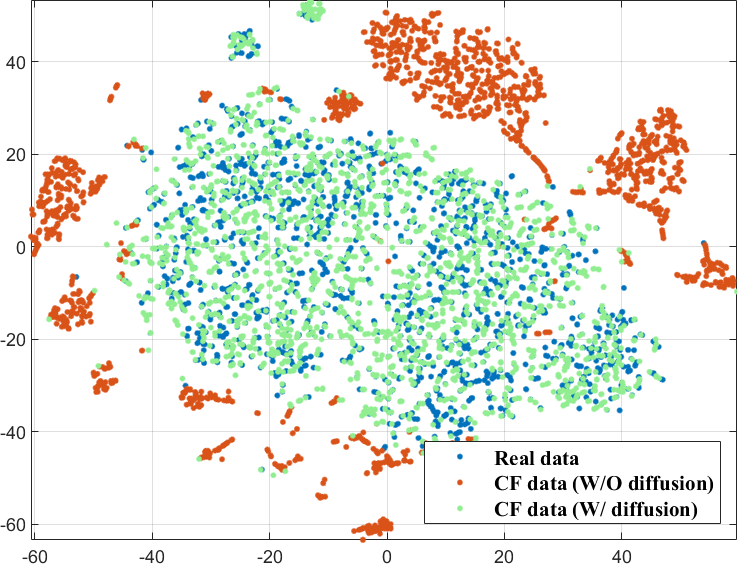}
\caption{T-SNE visualization of real data (blue), counterfactual data generated without diffusion (red), and counterfactual data generated with diffusion (light green).}
\label{tsne}
\end{figure}

Moreover, the t-SNE visualization, constructed from the data of all cases, qualitatively reinforces these findings. As shown in Fig. \ref{tsne}, the real data points (blue) form a cohesive central distribution. The counterfactual samples generated with diffusion refinement (light green) are densely interwoven with the real data, reflecting strong semantic alignment and faithful adherence to the underlying data manifold. In contrast, the counterfactuals produced without diffusion refinement (red) tend to scatter farther from the real data cluster, indicating a pronounced deviation from the authentic data distribution. This clear visual separation highlights the crucial role of the refinement module in producing anatomically coherent and semantically consistent counterfactuals.

Overall, these results underscore the importance of the diffusion refinement process in producing high-quality, clinically coherent counterfactuals.


\subsection{Interpretable latent space dynamics}
\label{T}
Our quantitative analysis of transition trajectories governed by the SDE, as illustrated in Fig. \ref{box}, reveals that the learned latent space exhibits a topology closely aligned with the clinical understanding of KOA progression. Specifically, we introduce the minimum required simulation time, $T_{\text{SDE}}^{\text{min}}$, as a surrogate measure for the energy barrier between adjacent KL grades. To quantify this barrier, we defined a successful transition as the point at which the classifier's output probability for the target KL grade surpasses the 0.5 threshold. It is noteworthy that the classifier's boundary in the latent space is not typically a broad, uniform region but rather can be conceptualized as a narrow and steep cliff, which implies that a very small increment in the simulation time $T_{\text{SDE}}$ can cause the generated sample's classification probability to jump abruptly, often from below 0.5 to 0.8 or higher. To precisely determine this threshold, we employed a binary search algorithm, performing 100 iterations per input image to accurately locate the $T_{\text{SDE}}^{\text{min}}$ at which this sharp transition occurs. This methodology reveals two fundamental properties of the learned manifold. Specifically:

\begin{figure}[htbp]
\centering
\includegraphics[width=0.4\textwidth]{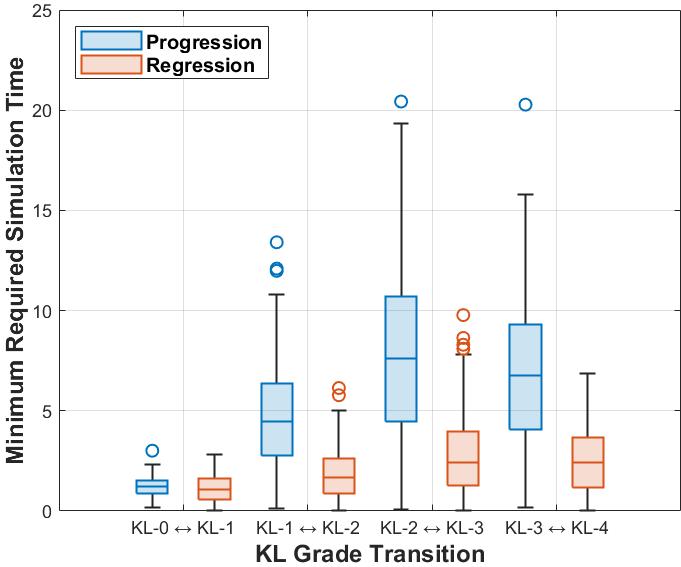}
\caption{Quantitative analysis of the computational effort for transitions between adjacent KL grades. The box plot illustrates the minimum required simulation time $T_{\text{SDE}}^{\text{min}}$ for KOA progression and regression.}
\label{box}
\end{figure}

\subsubsection{Progression} First, the progression analysis demonstrates a non-linear relationship between disease severity and the computational effort required for progression. Rather than a simple monotonic increase, the difficulty of traversing to the next KL grade peaks at the KL-2 to KL-3 transition, which demands the greatest mean $T_{\text{SDE}}^{\text{min}}$. The difficulty for the subsequent KL-3 to KL-4 transition, while still substantial, is relatively lower, which suggests the transition from KL-2 to KL-3 represents the most profound pathological change, involving the formation of significant JSN.

\subsubsection{Regression} Second, the regression analysis reveals a clear asymmetry in transition difficulty. Across all examined KL grade pairs, the mean $T_{\text{SDE}}^{\text{min}}$ for regressive transitions is consistently and significantly lower than for their progressive counterparts. Such an asymmetry offers interpretable insight into the model’s generative process. Specifically, regressive transitions primarily involve reductive operations, such as smoothing or removing pathological features, which are computationally simpler. In contrast, progressive transitions necessitate the synthesis of novel and pathologically meaningful structures, which imposes a higher computational burden.

Overall, $T_{\text{SDE}}^{\text{min}}$ serves as a principled and interpretable metric for probing the structure of the learned latent space that captures the non-linear, asymmetric, and progressive nature of KOA development, offering a topology that faithfully mirrors clinically observed patterns of disease advancement.

\subsection{Comparisons with state-of-the-art methods}
To comprehensively evaluate the effectiveness of our proposed method, we conduct comparisons against existing State-Of-The-Art (SOTA) approaches. The comparison is divided into two main parts: Qualitative visualization analysis and quantitative metric-based analysis. It is noteworthy that all shown generated samples in this section were deemed valid only if the frozen reference classifier assigned them to the intended KL grade with confidences exceeding 0.95, thereby ensuring semantic plausibility.

\subsubsection{Qualitative visualization analysis}
To facilitate a focused qualitative assessment, Table \ref{visualization} provides a visualization comparison of our proposed approach against three representative SOTA methods: DiverseCF \cite{mothilal2020explaining} (optimization-based), GANterfactual \cite{mertes2022ganterfactual} (GAN-based), and DiME \cite{jeanneret2022diffusion} (diffusion-based). To ensure a fair and direct comparison, all methods were evaluated on their ability to perform mutual counterfactual conversions between KL-2 and KL-3 (i.e., KL-2 $\Leftrightarrow$ KL-3), as demonstrated in our analysis in Section \ref{T}, the potential barrier for conversion between KL-2 and KL-3 is the highest among all adjacent grades, making it the most indicative case for evaluating model performance. All evaluated models were trained from scratch on the same dataset and leveraged the same frozen reference classifier from \cite{wang2024transformer} to guide the generation process. As depicted in Table \ref{visualization}, our method demonstrates a superior ability to generate high-fidelity and clinically plausible counterfactuals. For the progression task (KL-2 $\rightarrow$ KL-3), our model successfully introduces key pathological features (i.e., JSN) that are characteristic of KL-3. In contrast, while GANterfactual and DiME produce plausible images, the generated changes are less pronounced and distinct. On the other hand, for the regression task (KL-3 $\leftarrow$ KL-2), our method removes osteophytes to generate an image that conforms to KL-2 while the other methods struggle to produce such clear and meaningful changes. It is noteworthy that despite the shown generated samples being subject to a validity constraint, deep learning models often function as "black boxes," offering little insight into the specific structural features that drive their decisions. As a result, without interpretable visualizations, one cannot be certain that the generative model is responding to clinically meaningful anatomical changes. In this case, the generated alterations might simply constitute imperceptible noise patterns that exploit the classifier's vulnerabilities, functionally similar to adversarial attacks, causing it to change its prediction rather than meaningful counterfactual transitions. Therefore, the qualitative visualization analysis underscores the robustness and effectiveness of our proposed approach compared to existing SOTA methods.

\begin{table*}[htbp]
\centering
\caption{Qualitative visualization comparison with SOTA methods}
\label{visualization}
\begin{threeparttable}
\setlength{\tabcolsep}{0.2mm}
\begin{tabular}{cccccc}
\toprule
&\multirow{2.5}*{Original starting image ($x$)}&\multicolumn{4}{c}{Target counterfactual image ($x^\prime$)}\\
\cmidrule(lr){3-6}
& & DiverseCF & GANterfactual & DiME & Ours\\
\midrule
\rotatebox[origin=c]{90}{\bf{KL-2 $\rightarrow$ KL-3}}&\begin{minipage}[b]{0.4\columnwidth}\centering \raisebox{-.5\height}{\includegraphics[width=\linewidth]{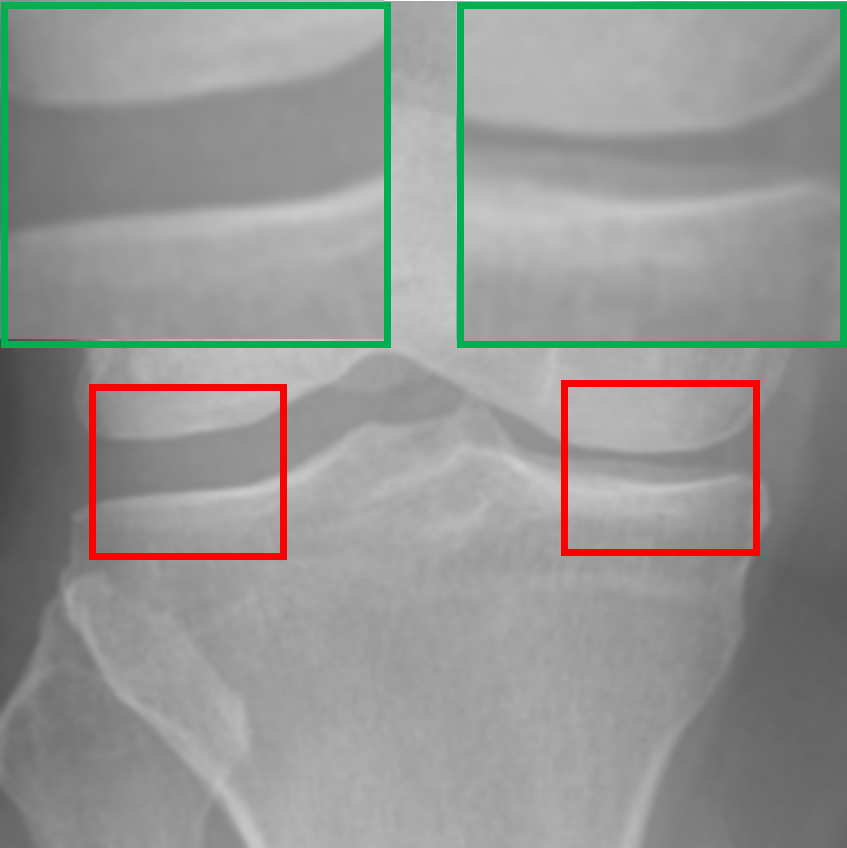}}\end{minipage}&\begin{minipage}[b]{0.4\columnwidth}\centering \raisebox{-.5\height}{\includegraphics[width=\linewidth]{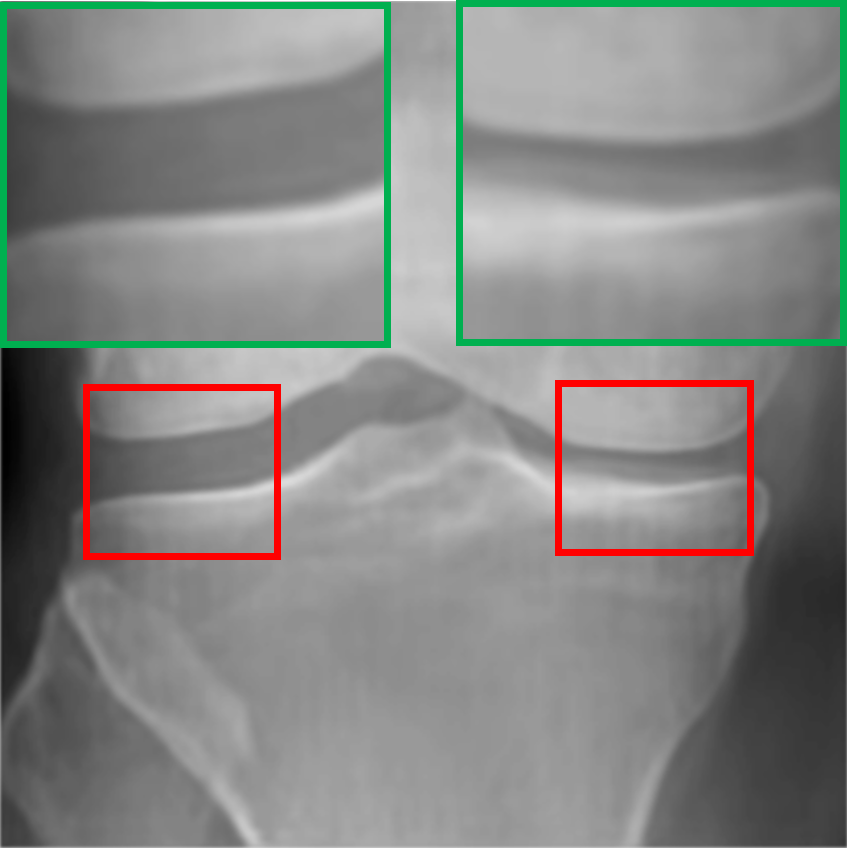}}\end{minipage}&\begin{minipage}[b]{0.4\columnwidth}\centering \raisebox{-.5\height}{\includegraphics[width=\linewidth]{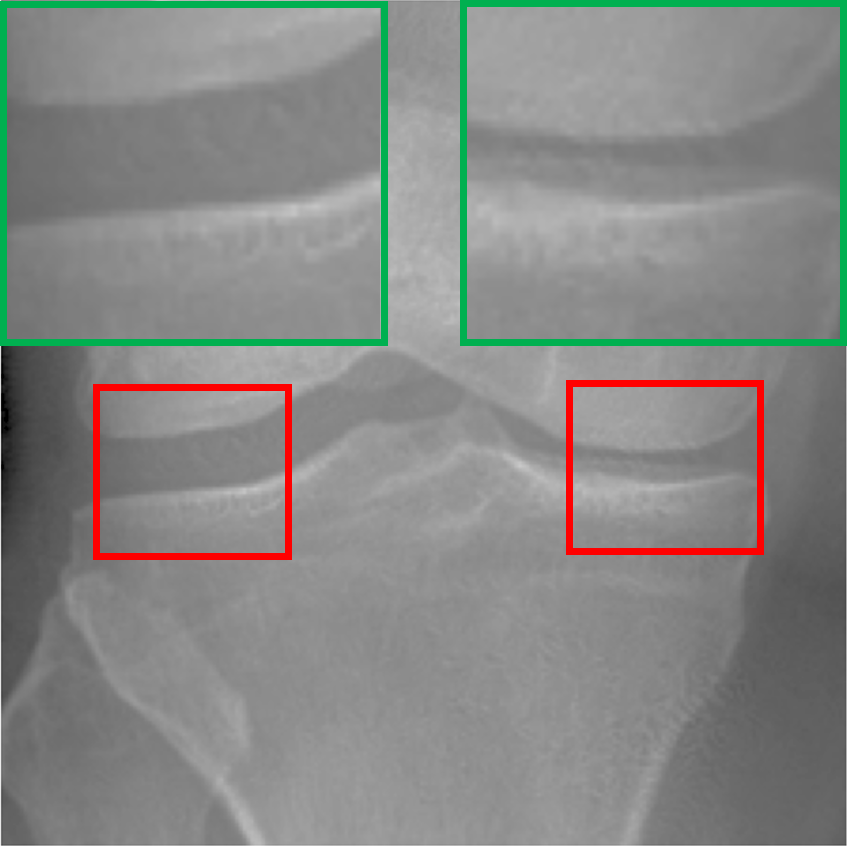}}\end{minipage}&\begin{minipage}[b]{0.4\columnwidth}\centering \raisebox{-.5\height}{\includegraphics[width=\linewidth]{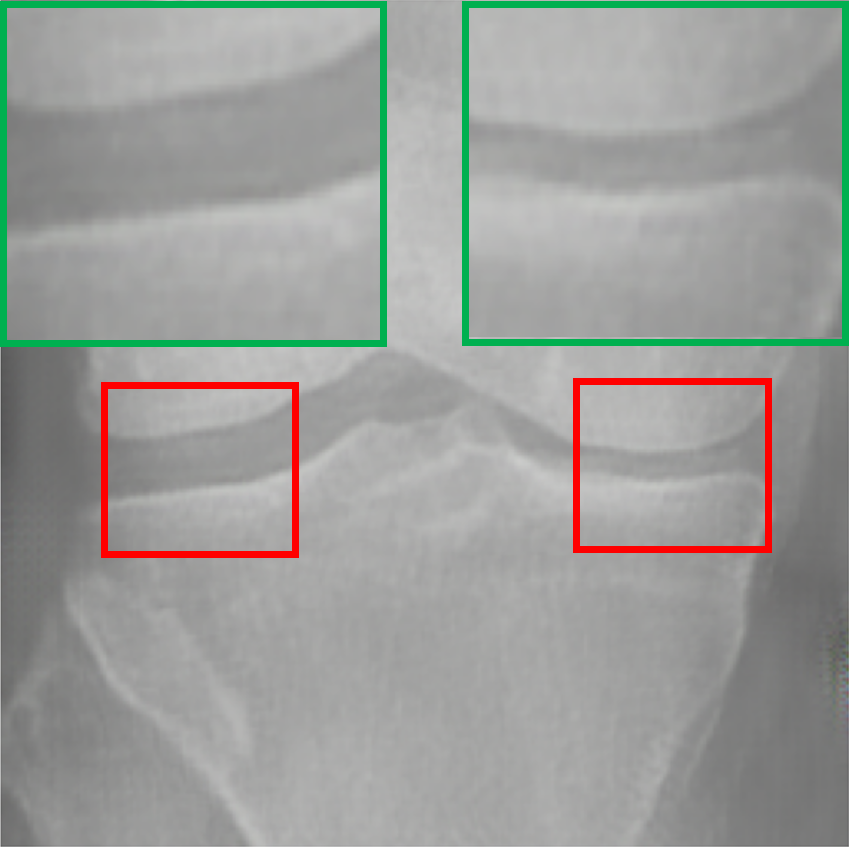}}\end{minipage}&\begin{minipage}[b]{0.4\columnwidth}\centering \raisebox{-.5\height}{\includegraphics[width=\linewidth]{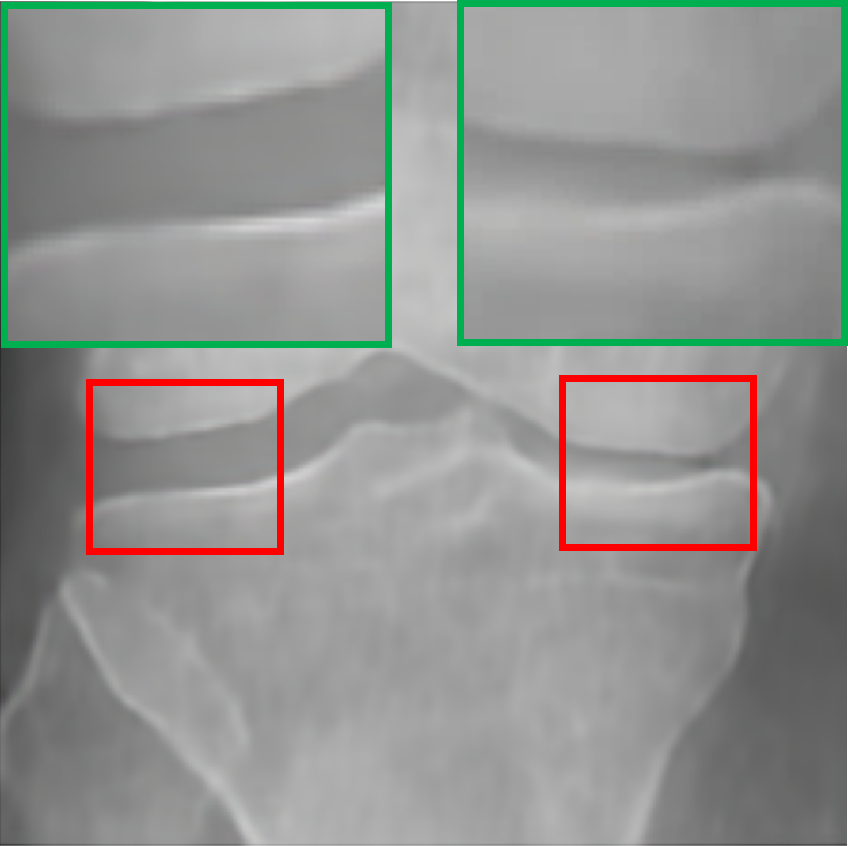}}\end{minipage}\\
\midrule
\rotatebox[origin=c]{90}{\bf{KL-2 $\leftarrow$ KL-3}}&\begin{minipage}[b]{0.4\columnwidth}\centering \raisebox{-.5\height}{\includegraphics[width=\linewidth]{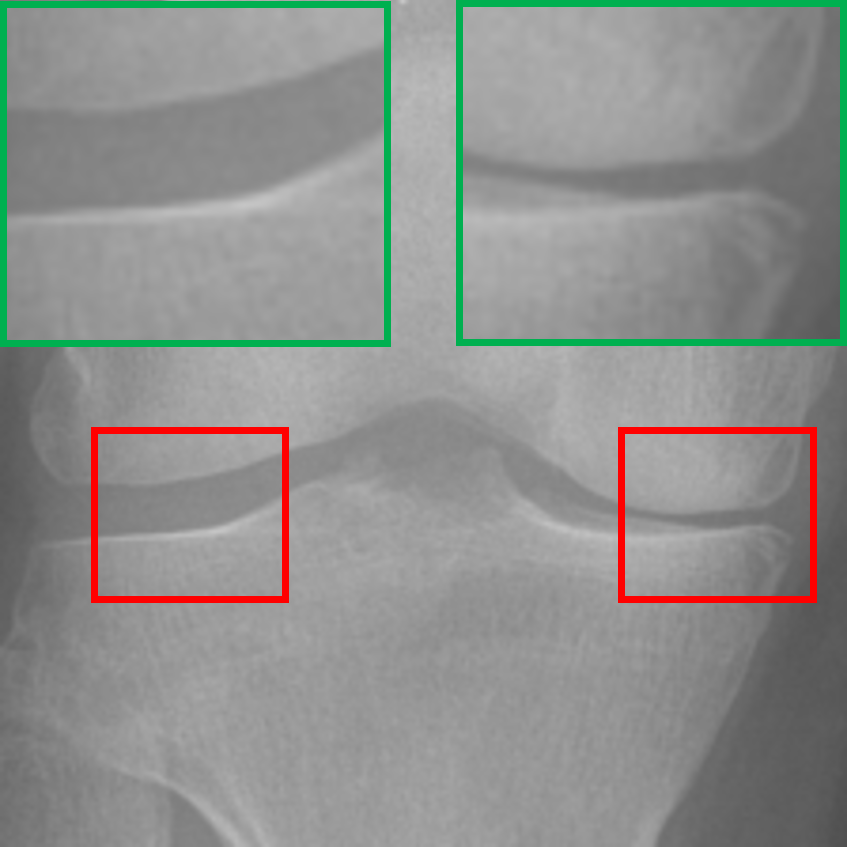}}\end{minipage}&\begin{minipage}[b]{0.4\columnwidth}\centering \raisebox{-.5\height}{\includegraphics[width=\linewidth]{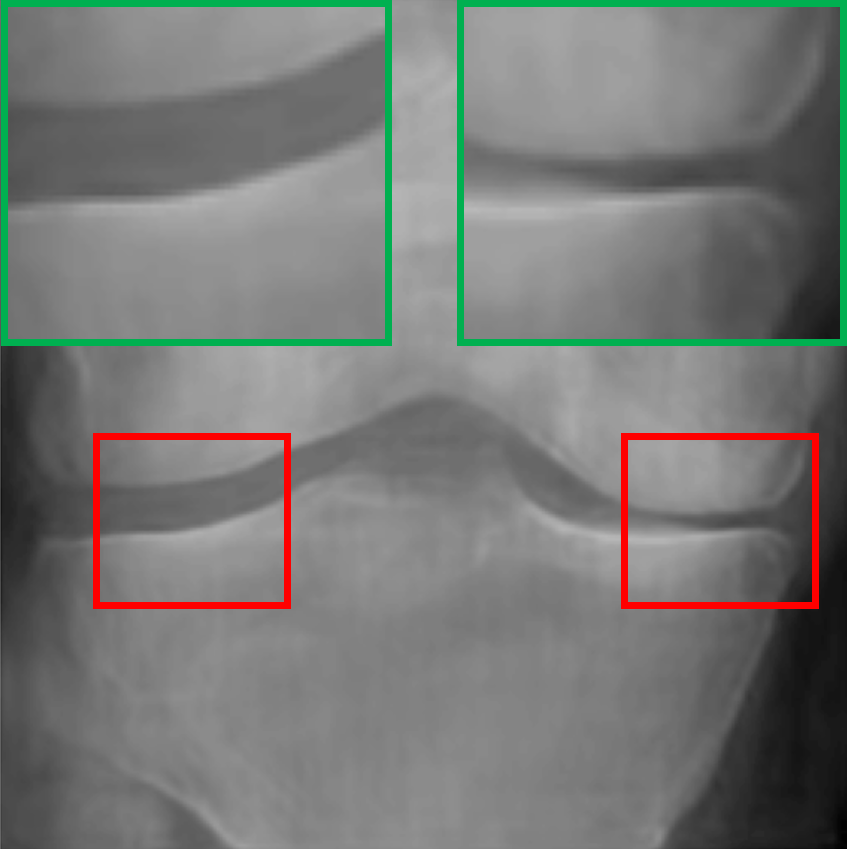}}\end{minipage}&\begin{minipage}[b]{0.4\columnwidth}\centering \raisebox{-.5\height}{\includegraphics[width=\linewidth]{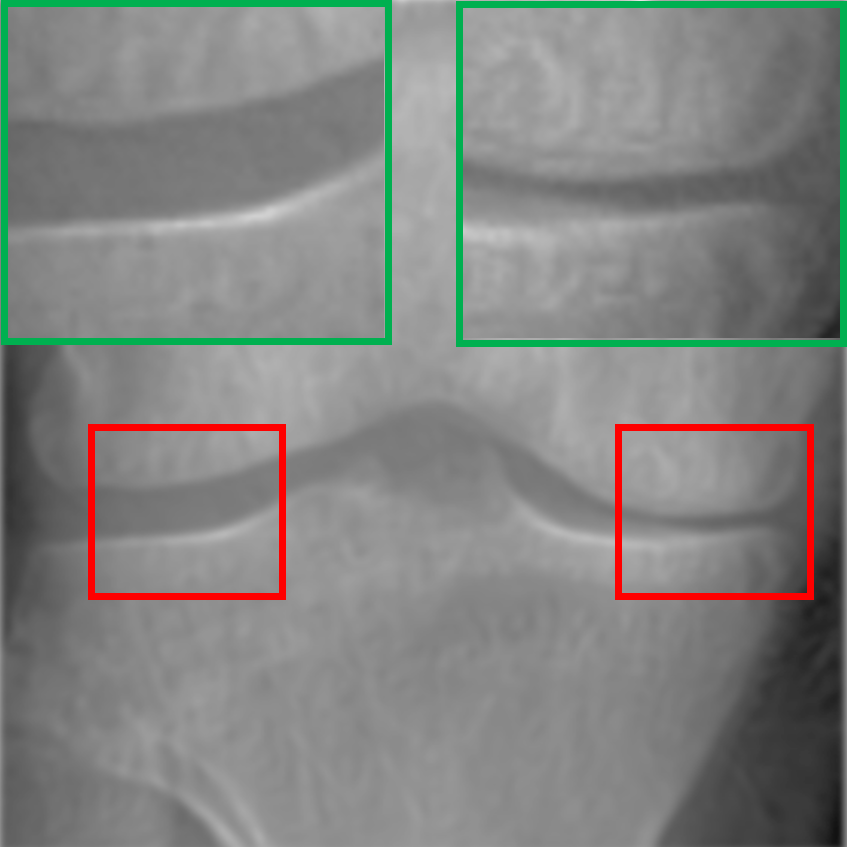}}\end{minipage}&\begin{minipage}[b]{0.4\columnwidth}\centering \raisebox{-.5\height}{\includegraphics[width=\linewidth]{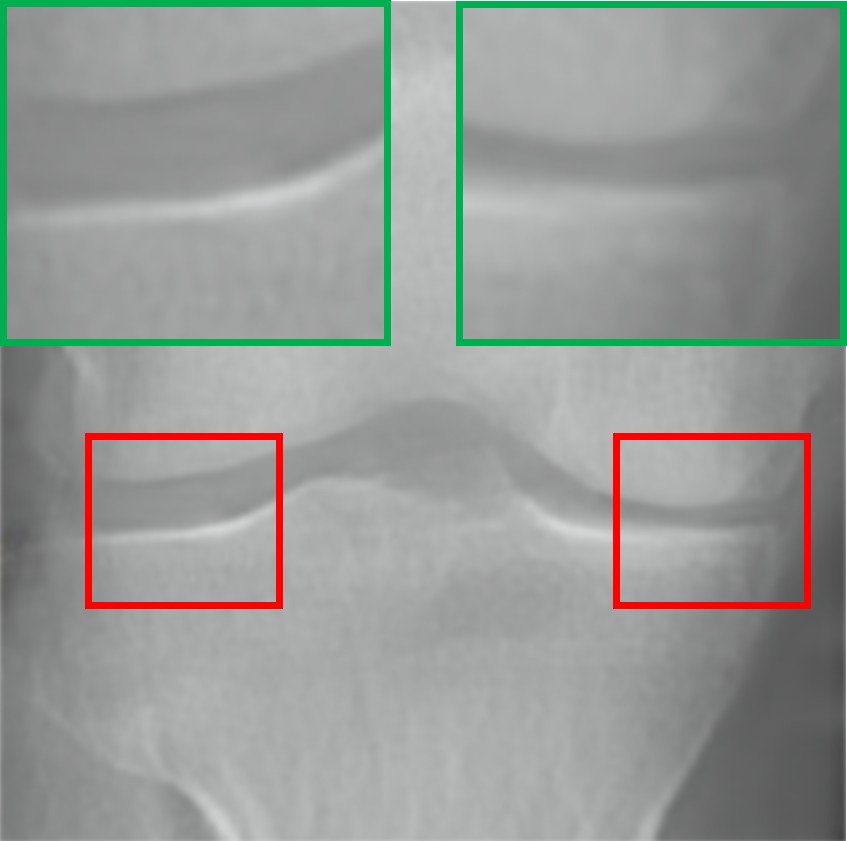}}\end{minipage}&\begin{minipage}[b]{0.4\columnwidth}\centering \raisebox{-.5\height}{\includegraphics[width=\linewidth]{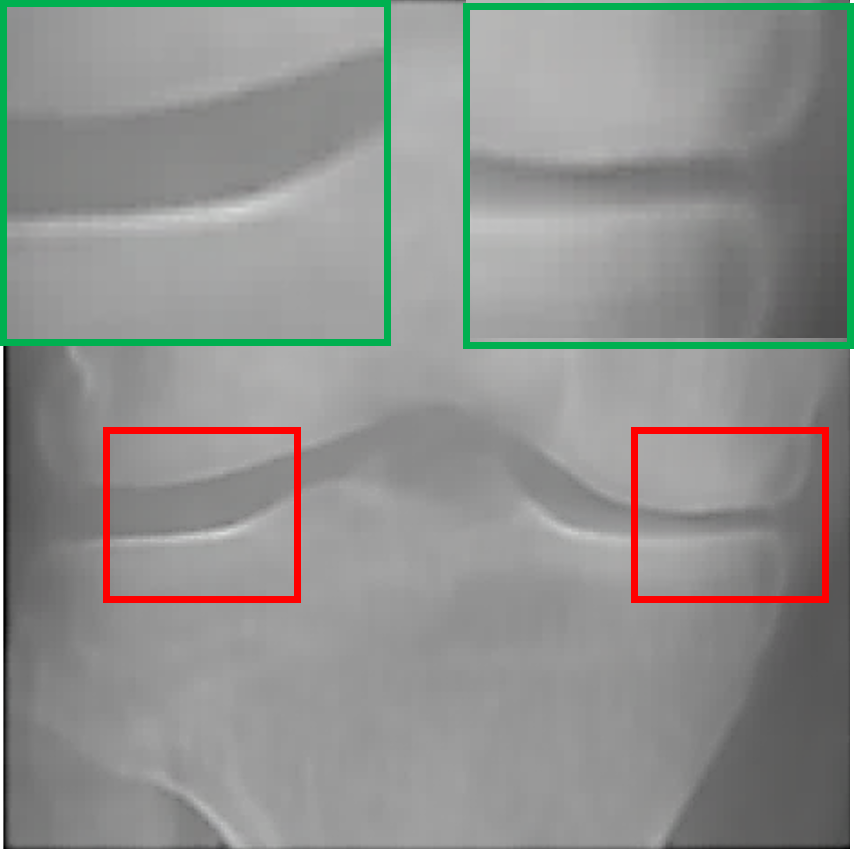}}\end{minipage}\\
\bottomrule
\end{tabular}
\end{threeparttable}
\end{table*}

\begin{table}[htbp]
\centering
\caption{Quantitative comparison across SOTA methods}
\label{metrics}
\begin{threeparttable}
\setlength{\tabcolsep}{0.3mm} 
\begin{tabular}{c|lcccc}
\toprule
 &Method& PSNR  & SSIM  & LPIPS & RMSE \\
\hline
\multirow{5}*{\rotatebox[origin=c]{90}{\bf{0 $\leftrightarrow$ 1}}} &Ref & $29.03, 29.31$ & $0.881, 0.848$ & $4.324, 4.211$ & $14.68, 16.25$\\
\cline{2-6}
&DiverseCF & $30.15$ / $30.62$ & $0.885$ / $0.843$ & $3.148$ / $3.141$ & $6.81$ / $6.22$\\
&GANterfactual & $32.02$ / $32.58$ & $0.896$ / $0.865$ & $2.121$ / $2.115$ & $3.75$ / $3.11$\\
&DiME & $33.49$ / $34.05$ & $0.913$ / $0.891$ & $1.098$ / $1.092$ & $1.82$ / $1.21$\\
&Ours & \bf $35.51$ / $36.23$ & \bf $0.925$ / $0.904$ & \bf $1.085$ / $1.080$ & \bf $1.12$ / $1.07$ \\
\hline
\multirow{5}*{\rotatebox[origin=c]{90}{\bf{1 $\leftrightarrow$ 2}}} &Ref & $29.31, 29.03$ & $0.848, 0.826$ & $4.211, 5.651$ & $16.25, 20.89$ \\
\cline{2-6}
&DiverseCF & $29.88$ / $30.41$ & $0.851$ / $0.849$ & $3.153$ / $3.145$ & $9.22$ / $9.58$\\
&GANterfactual & $31.75$ / $32.33$ & $0.879$ / $0.873$ & $2.129$ / $2.123$ & $6.15$ / $6.55$\\
&DiME & $33.15$ / $33.81$ & $0.886$ / $0.885$ & $1.105$ / $1.072$ & $2.21$ / $2.59$\\
&Ours & \bf $34.17$ / $34.85$ & \bf $0.907$ / $0.899$ & \bf $1.095$ / $1.060$ & \bf $1.93$ / $1.28$ \\
\hline
\multirow{5}*{\rotatebox[origin=c]{90}{\bf{2 $\leftrightarrow$ 3}}} &Ref & $29.03, 26.84$ & $0.826, 0.802$ & $5.651, 9.421$ & $20.89, 27.55$ \\
\cline{2-6}
&DiverseCF & $29.35$ / $26.88$ & $0.831$ / $0.829$ & $4.185$ / $5.178$ & $10.55$ / $10.96$\\
&GANterfactual & $29.85$ / $28.41$ & $0.853$ / $0.842$ & $3.156$ / $4.150$ & $7.43$ / $7.85$\\
&DiME & $30.88$ / $30.45$ & $0.870$ / $0.866$ & $2.132$ / $2.126$ & $3.33$ / $3.78$\\
&Ours & \bf $32.45$ / $31.18$ & \bf $0.882$ / $0.883$ & \bf $1.112$ / $1.507$ & \bf $2.82$ / $2.31$ \\
\hline
\multirow{5}*{\rotatebox[origin=c]{90}{\bf{3 $\leftrightarrow$ 4}}} &Ref & $26.84, 25.11$ & $0.802, 0.759$ & $9.421, 11.823$ & $27.55, 39.27$\\
\cline{2-6}
&DiverseCF & $27.65$ / $25.17$ & $0.828$ / $0.787$ & $6.178$ / $8.171$ & $10.58$ / $10.99$\\
&GANterfactual & $29.55$ / $28.15$ & $0.845$ / $0.857$ & $4.154$ / $5.148$ & $8.48$ / $8.87$\\
&DiME & $30.01$ / $30.63$ & $0.861$ / $0.861$ & $3.129$ / $3.123$ & $4.41$ / $4.78$\\
&Ours & \bf $31.83$ / $31.52$ & \bf $0.885$ / $0.883$ & \bf $1.703$ / $2.098$ & \bf $2.94$ / $2.37$ \\
\bottomrule
\end{tabular}
\begin{tablenotes}
\footnotesize
\item[$*$] Ref denotes the reference baseline, quantifying the inherent intra-class variability, with the values on the left/right of the comma (,) representing the average dissimilarity within the lower and higher KL grade image sets, respectively.
\item[$**$] The values on the left/right of the slash (/) are the metrics computed between the generated counterfactual images and their original images, corresponding to the progression and regression directions, respectively.
\end{tablenotes}
\end{threeparttable}
\end{table}

\subsubsection{Quantitative metric-based analysis}
To facilitate a robust and equitable quantitative comparison, we generated ten independent counterfactual samples for each input image using every evaluated method and computed the average across multiple performance metrics. Table \ref{metrics} reports the results across standard image quality metrics, including PSNR (dB), SSIM, LPIPS ($10^{-2}$), and RMSE ($10^{-2}$). As can be seen, all evaluated methods significantly outperform the reference baseline, producing higher-fidelity outputs. Furthermore, our proposed method consistently achieves superior performance across all KL-grade transitions. A clear trend emerges in which performance declines as the transition task becomes more difficult, reflecting the increased anatomical complexity of advanced KOA stages. However, our method demonstrates stronger robustness to this degradation. For instance, PSNR decreases by only around 4 dB from the easiest transitions (KL-0 $\rightarrow$ KL-1: 35.51 dB; KL-1 $\rightarrow$ KL-0: 36.23 dB) to the most challenging ones (KL-3 $\rightarrow$ KL-4: 31.83 dB; KL-4 $\rightarrow$ KL-3: 31.52 dB), indicating a relatively stable generative performance under increasing structural variation. Among the SOTA approaches, DiME, a diffusion-based method, performs competitively and consistently ranks second across all metrics, although it trails slightly behind our model. GANterfactual, a GAN-based technique, achieves moderate performance, but its quality diminishes notably for higher-grade transitions, possibly due to its limited ability to preserve fine structural details under large domain shifts. On the other hand, DiverseCF, an optimization-based method that directly perturbs pixel values without leveraging a generative prior, performs the least effectively, underscoring the challenges of maintaining high perceptual quality in the absence of strong structural constraints.

Overall, our approach demonstrates both stable and high-fidelity performance in progression and regression directions, reflecting its robust and generalizable capability for clinically meaningful counterfactual generation.





%
%
\begin{table*}[htbp]
\centering
\caption{Counterfactual augmentation for adjacent KL grade classification on OAI and MOST}
\setlength{\tabcolsep}{0.9mm} 
\begin{threeparttable}
\begin{tabular}{l|lccc|ccc|c|lcc|ccc} 
\toprule
&\multirow{2.2}*{Model} & \multicolumn{3}{c|}{OAI} & \multicolumn{3}{c|}{MOST}&& \multicolumn{3}{c|}{OAI} & \multicolumn{3}{c}{MOST}\\
\cline{3-8} \cline{10-15}
& & Acc& Acc$^\diamond$ &$p$-value & Acc & Acc$^\diamond$ &$p$-value&& Acc & Acc$^\diamond$ &$p$-value& Acc & Acc$^\diamond$ &$p$-value\\
\hline
\multirow{6}*{\rotatebox[origin=c]{90}{\bf{KL-0 vs. KL-1}}} & VGG-11 & $60.35$ & $60.92$ & $1.82e-01$ & $58.72$ & $58.65$ & $2.15e-01$ &\multirow{6}*{\rotatebox[origin=c]{90}{\bf{KL-1 vs. KL-2}}} & $68.39$ & $71.58$ & $1.65e-04^{***}$ & $67.21$ & $68.33$ & $5.95e-03^{**}$ \\
&EfficientNet-B0 & $62.48$ & $62.97$ & $1.53e-02^{*}$ & $60.91$ & $61.33$ & $1.78e-01$ && $69.88$ & $72.03$ & $1.25e-03^{**}$ & $68.59$ & $71.88$ & $1.53e-03^{**}$ \\
&Inception-V3 & $63.89$ & $64.05$ & $1.21e-01$ & $61.55$ & $62.12$ & $1.49e-02^{*}$ && $70.53$ & $72.88$ & $9.95e-03^{**}$ & $69.21$ & $71.59$ & $8.20e-04^{***}$ \\
&ViT-B/16 & $64.52$ & $65.99$ & $4.82e-02^{*}$ & $62.33$ & $63.01$ & $1.15e-02^{*}$ && $71.28$ & $74.53$ & $7.80e-05^{***}$ & $70.03$ & $72.45$ & $9.28e-03^{**}$ \\
&Tiulpin et al. \cite{tiuplin} & $65.88$ & $66.32$ & $2.51e-02^{*}$ & $63.78$ & $64.21$ & $9.90e-02$ && $74.33$ & $76.67$ & $8.22e-04^{***}$ & $69.11$ & $71.53$ & $1.05e-03^{**}$ \\
&Wang et al. \cite{wang2024transformer} & $69.08$ & $69.11$ & $4.33e-01$ & $66.57$ & $66.95$ & $5.54e-02$ && $77.02$ & $79.11$ & $9.18e-04^{***}$ & $75.88$ & $77.21$ & $1.22e-04^{***}$ \\
\hline
\multirow{6}*{\rotatebox[origin=c]{90}{\bf{KL-2 vs. KL-3}}} & VGG-11 & $84.67$ & $86.33$ & $5.66e-04^{***}$ & $83.98$ & $85.58$ & $1.91e-03^{**}$ &\multirow{6}*{\rotatebox[origin=c]{90}{\bf{KL-3 vs. KL-4}}} & $86.33$ & $88.02$ & $1.73e-04^{***}$ & $84.67$ & $85.38$ & $2.11e-04^{***}$ \\
&EfficientNet-B0 & $86.79$ & $87.21$ & $1.20e-03^{**}$ & $85.88$ & $87.43$ & $1.45e-03^{**}$ && $86.69$ & $88.12$ & $2.33e-03^{**}$ & $86.93$ & $87.58$ & $6.55e-04^{***}$ \\
&Inception-V3 & $87.83$ & $88.53$ & $9.91e-03^{**}$ & $87.03$ & $87.78$ & $4.14e-04^{***}$ && $87.88$ & $88.33$ & $9.58e-03^{**}$ & $88.12$ & $88.78$ & $1.15e-02^{*}$ \\
&ViT-B/16 & $88.59$ & $90.38$ & $7.54e-03^{**}$ & $87.92$ & $88.77$ & $8.89e-04^{***}$ && $88.29$ & $89.03$ & $7.16e-03^{**}$ & $88.93$ & $90.58$ & $8.51e-03^{**}$ \\
&Tiulpin et al. \cite{tiuplin} & $90.87$ & $91.58$ & $2.39e-03^{**}$ & $89.03$ & $90.99$ & $2.90e-03^{**}$ && $90.13$ & $90.68$ & $2.50e-03^{**}$ & $90.53$ & $91.33$ & $3.21e-03^{**}$ \\
&Wang et al. \cite{wang2024transformer} & $91.23$ & $91.98$ & $2.16e-03^{**}$ & $90.02$ & $90.87$ & $1.55e-03^{**}$ && $90.72$ & $91.63$ & $8.82e-03^{**}$ & $91.21$ & $91.94$ & $1.18e-03^{**}$ \\
\bottomrule
\end{tabular}
\begin{tablenotes}
\footnotesize
\item[$\diamond$] Accuracy obtained after the data augmentation based on self-corrective learning.
\item[$*$] $p$-values were computed using an independent t-test. Statistical significance is indicated as follows: $^{***}$$p$ $<$ 0.001, $^{**}$$p$ $<$ 0.01, $^{*}$$p$ $<$ 0.05.
\end{tablenotes}
\end{threeparttable}
\label{comparision_adjacent}
\end{table*}

\subsection{Effect of self-corrective learning}
We designed a targeted data augmentation strategy to enhance the classifier's robustness by leveraging the empirically determined transition barriers to generate high-quality and boundary-informed counterfactual samples. To validate the effect of self-correction, each experiment adhered to a closed-loop design that the same classifier architecture used to define decision boundaries for counterfactual generation was subsequently fine-tuned on the generated samples and re-evaluated. The core idea of our approach is to assign an appropriate simulation time $T_\text{SDE}$ for each specific transition between adjacent KL grades. The data generation process was bidirectional, for each pair of adjacent KL grades, the counterfactual samples were generated in both the progression and regression directions. It is noteworthy that this focus on adjacent pairs is methodologically critical as transitions between non-adjacent grades are inherently ambiguous, and crossing a classifier’s decision boundary in such cases may simply indicate an intermediate or spurious state rather than a valid mapping to the intended target grade. To ensure that generated samples confidently reside within the target class manifold, we set the simulation time for each transition to exceed its average minimum barrier. Specifically, the simulation time was defined as $T_{\text{SDE}} = 1.5 \times \bar{T}_{\text{SDE}}^{\text{min}}$, where $\bar{T}_{\text{SDE}}^{\text{min}}$ is the average minimum transition time computed in Section \ref{T}. The scaling factor of 1.5 was selected empirically to guarantee that samples not only cross the boundary but also stabilize within the semantic core of the target distribution, providing strong and unambiguous supervision for training. During each 1000-step SDE simulation, an intermediate sample was extracted every 100 steps, yielding 10 distinct augmented samples per original image. Each sample’s label was dynamically assigned based on the classifier’s instantaneous probability distribution at the corresponding simulation step. These counterfactual samples were added exclusively to the original training set, and all experiments were repeated independently three times. The final reported metrics represent the average performance across these trials.

The results detailed in Table \ref{comparision_adjacent} show a broad trend of consistent and often statistically significant improvements in classification accuracy across various models and tasks on both the OAI and MOST datasets, with notable enhancements observed even in tasks with high baseline performance, such as KL-2 vs. KL-3 and KL-3 vs. KL-4. Moreover, a particularly illustrative case arises from the behaviour surrounding the KL-1 grade, which is clinically defined by "doubtful" pathological signs and is thus subject to significant semantic ambiguity. Specifically, for the KL-0 vs. KL-1 classification, the proposed augmentation strategy yields only marginal and often statistically insignificant performance gains. In contrast, for KL-1 vs. KL-2, the same approach consistently produces substantial and statistically robust improvements across all evaluated architectures, which likely reflects a latent-space topology in which the KL-1 manifold lies closer to KL-0 than to KL-2, making the KL-0 vs. KL-1 boundary inherently less well-defined and more resistant to augmentation-based improvement. Such interpretation is further supported by the energy barrier analysis in Section \ref{T}, where the average minimum simulation time ($T_{\text{SDE}}^{\text{min}}$) required to traverse from KL-0 to KL-1 is lower than that for KL-1 to KL-2. These findings collectively highlight the sensitivity of our self-corrective augmentation framework to the underlying semantic and topological structure of the KOA progression continuum. These findings collectively demonstrate the potential of self-corrective learning to enhance classifier robustness through targeted counterfactual samples, while also highlighting its sensitivity to the underlying semantic and topological structure of the KOA progression continuum.

\section{Conclusion and Discussion}
In this study, we proposed Diffusion-based Counterfactual Augmentation (DCA), a novel framework aimed at improving both the robustness and interpretability of models for KL grading in Knee OsteoArthritis (KOA). The framework introduces a gradient-informed counterfactual generation process that explicitly targets a classifier’s decision boundaries, enabling the synthesis of clinically meaningful augmented samples that expose regions of model uncertainty. By leveraging a Stochastic Differential Equation (SDE) to explore transitions in the latent space between adjacent KL grades, DCA produces high-fidelity counterfactuals that are further refined via a diffusion-based denoising process to ensure adherence to the true data manifold. These refined samples are incorporated into a self-corrective learning loop, wherein a frozen reference classifier identifies boundary-sensitive regions and a learnable target classifier is retrained on the generated samples to resolve local ambiguities. Extensive experiments validate the effectiveness of DCA. Compared to State-Of-The-Art (SOTA) methods, our approach yields superior image quality and clinical plausibility. Furthermore, integrating DCA-generated samples into training pipelines consistently led to statistically significant improvements in classification accuracy across diverse architectures on both the OAI and MOST datasets.

Several key aspects of this study warrant further in-depth discussion.

\textbf{The self-correction paradox: Can a model teach itself?} A fundamental question emerges from our data augmentation strategy: how can a model improve its own performance by training on samples generated using its own, presumably imperfect, decision boundary? The resolution lies in the model’s ability to convert uncertainty into certainty through a principled refinement process. Rather than simply reinforcing its existing beliefs, our method establishes a structured self-corrective loop. Specifically, it begins by leveraging the classifier’s own gradient to identify the most efficient trajectory for altering a sample’s predicted class. Then, by applying a simulation time, the framework pushes an initially ambiguous, low-confidence boundary case deep into the semantic territory of the target class, yielding a high-confidence counterfactual. When this new, definitive sample is reintroduced into the training set, it acts as a stronger and more definitive supervisory signal than the original borderline case could provide. This targeted augmentation compels the model to refine its internal representations and construct more resilient decision boundaries. In essence, the process forms a closed feedback loop where latent uncertainty is externalized, transformed into confident knowledge, and re-learned, enabling the model to systematically correct the very regions where it was previously most uncertain.

\textbf{Rethinking of clinical ambiguity:} The tempered performance of our framework on the clinically ambiguous KL-0 vs. KL-1 task should not be seen as a limitation, yet as evidence of its sophisticated sensitivity to the underlying data topology. KL-1 denotes “doubtful” pathological signs, occupying a semantic grey zone where even expert human raters exhibit low inter-rater agreement. Rather than fabricating artificial boundaries, DCA respects this ambiguity, enhancing learning where meaningful variation exists and exercising restraint where it does not. Such behaviour reflects a trustworthy and interpretable AI system, one that supports clinical reasoning without overstating its certainty.

\textbf{Strengths:} A central strength of this work lies in its novel counterfactual generation framework, which explicitly targets a classifier’s decision boundaries to improve model robustness. Our approach employs a gradient-informed boundary drive to steer samples into semantically ambiguous regions, while a manifold constraint maintains anatomical and clinical plausibility. As demonstrated through both quantitative metrics and qualitative visualizations, this design yields counterfactuals with superior reconstruction fidelity and diagnostic relevance. Moreover, this study also involves a multi-faceted interpretability discussion, which offers not only intuitive visualization of minimal anatomical changes responsible for grade transitions but also introduces a quantitative interpretability metric, the minimum required simulation time, which serves as a principled probe into the latent disease manifold. This metric reveals a meaningful and clinically aligned topological structure of KOA progression, reflecting the non-linear, asymmetric nature of pathological transitions and offering deeper insight into how the model internalizes disease severity.

\textbf{Limitations:} Despite its demonstrated strengths, the proposed framework has several limitations inherent in its current design and scope. Firstly, the methodology is intentionally focused on generating counterfactuals exclusively between adjacent KL grades, thus the framework in its present form does not model direct, long-range disease progression. Second, the counterfactual generation process is entirely dependent on gradients derived from a frozen reference classifier. Consequently, the quality and clinical validity of the generated samples are heavily influenced by the performance and representational capacity of this pre-trained model. Finally, the process for generating samples that lie confidently within a target class manifold relies on an empirically determined hyperparameter.

\textbf{Future work:} Future work could focus on modelling the entire spectrum of KOA as a continuous pathological evolution, which would enable the generation of counterfactuals between any two disease states, providing a more holistic simulation of progression. To improve the framework’s autonomy and generalizability, another promising direction involves developing adaptive control mechanisms for determining the optimal SDE simulation time. Such mechanisms would eliminate reliance on fixed, empirically chosen scaling factors and instead dynamically estimate the appropriate simulation horizon required for successful class transitions. Finally, the applicability of this approach could be extended to other challenging medical imaging tasks that involve ordinal classification and ambiguous decision boundaries, such as diabetic retinopathy staging or Alzheimer’s disease classification, where similar issues of semantic continuity and boundary uncertainty persist.

\section{Statements}
This manuscript was prepared using data from OAI and MOST. The views expressed in it are those of the authors and do not necessarily reflect the opinions of the OAI and MOST investigators, the National Institutes of Health (NIH), or the private funding partners.

\section{Acknowledgements}
The authors gratefully acknowledge the funding from the Ralph Schlaeger Research Fellowship under Award No. 246448 at Massachusetts General Hospital (MGH), Harvard Medical School (HMS).

\bibliographystyle{IEEEtran}
\bibliography{references}

\end{document}